\documentclass{article}
\usepackage[a4paper, margin=1in]{geometry}
\usepackage{graphicx}
\usepackage[numbers]{natbib}
\usepackage{color}
\usepackage{tikz,pgfplots}
\usepackage{amsmath,amsfonts,amssymb}
\usepackage[symbol]{footmisc}
\usepackage{bm}
\usepackage{caption}
\usepackage{subcaption}
\usepackage[symbol]{footmisc}
\usepgfplotslibrary{external} 
\tikzset{external/up to date check={md5}}
\tikzset{external/mode=convert with system call} 
\begin{document}

\title{Stacked autoencoders based machine learning for noise reduction and signal reconstruction in geophysical data
}

\author{Debjani Bhowmick$^{1}$\footnotemark[2]
		\and Deepak K. Gupta$^{2}$\footnotemark[2] \and Saumen Maiti$^3$ \and Uma Shankar$^4$
}

\date{%
	$^1$Cognitive Science \& Artificial Intelligence, Tilburg University, The Netherlands \\
	$^2$Informatics Institute, University of Amsterdam, The Netherlands \\
	$^3$Dept. of Applied Geophysics, Indian Institute of Technology (ISM) Dhanbad, India \\
	$^4$Dept. of Geophysics, Banaras Hindu University, Varanasi, India \\ 
}

\footnotetext[2]{During this research, D. Bhowmick was affiliated with Indian Institute of Technology, ISM Dhanbad, India and Coal India Limited, Ranchi, India, and D. K. Gupta was affiliated with Delft University of Technology, The Netherlands.}
\maketitle

\begin{abstract}
Autoencoders are neural network formulations where the input and output of the network are identical and the goal is to identify the hidden representation in the provided datasets. Generally, autoencoders project the data nonlinearly onto a lower dimensional hidden space, where the important features get highlighted and interpretation of the data becomes easier. Recent studies have shown that even in the presence of noise in the input data, autoencoders can be trained to reconstruct the noisefree component of the data from the reduced-dimensional hidden space. 

In this paper, we explore the application of autoencoders within the scope of denoising geophysical datasets using a data-driven methodology. The autoencoder formulation is discussed, and a stacked variant of deep autoencoders is proposed. The proposed method involves locally training the weights first using basic autoencoders, each comprising a single hidden layer. Using these initialized weights as starting points in the optimization model, the full autoencoder network is then trained in the second step. The applicability of denoising autoencoders has been demonstrated on a basic mathematical example and several geophysical examples. For all the cases, autoencoders are found to significantly reduce the noise in the input data. 
\end{abstract}

\section{Introduction}
\label{intro}

Machine learning has been a trending topic in the past two decades, and it has widely been used in various science and engineering disciplines for improved interpretation of big datasets. Creating a machine learning algorithm essentially refers to building a model that can output approximately correct information when fed with certain input data. These models can be thought of as black-boxes: input goes in and output comes out - although the mapping from input to output can be fairly complex in itself. With the advent of powerful computers, it has become remarkably easy to train the computers to identify the hidden complex representations in provided datasets. For an overview of the applications of machine learning in various fields, see the recent review works presented in \cite{Tsai2009, Crisci2012, Kourou2015, Jones2016, Khan2018}, among others.  

Among the various machine learning methods, neural networks in particular, have received enormous attention. Here, we list some of the early works related to application of neural networks in various sectors. For a detailed overview, please see the citing references to these papers. In the finance sector, neural networks are used for bankruptcy prediction of banks/firms \cite{Salchenberger1992}, future options hedging and pricing \cite{Hutchinson1994}, credit evaluation \cite{Jenson1992}, interest rate prediction \cite{Nikolopoulos1994}, inter-market analysis \cite{RuggieroJr1994} and stock performance \cite{Bansal1993}. In human resources, the processes of personnel selection and workplace behavior prediction are automated using this technique \cite{Proctor1991, Collins1993}. In the information sector, neural networks have widely been used for authentication or identification of computer users \cite{Rogers1995}, recognition and classification of computer viruses \cite{Doumas1995}, pictorial information retrieval \cite{Stafylopatis1992}, \emph{etc}. Recently, neural networks have been used to solve several challenging problems in the field of medical imaging. For example, with this powerful tool, lung cancer can be effectively diagnosed and differentiated from lung benign diseases, normal control and gastrointestinal cancers \cite{Feng2012}. There is an unending list of other applications where neural networks have proved to be worthy, and not all of these can be listed here.  

The discipline of geophysics is no exception and neural networks have been used on various geophysical problems, \emph{e.g.}, inversion of electromagnetic, magnetelluric and seismic data \cite{Poulton1992, Zhang1997, Roth1994}, waveform recognition and first-break picking \cite{Murat1992}, trace editing \cite{McCormack1993}, lithological classification \cite{Maiti2007}, guide geophysical and geological modeling process \cite{Reading2015}, creating ensemble models for the estimation of petrophysical parameters \cite{Bhowmick2016}, \emph{etc}. 

The objective of most of the neural network based formulations is to mimic the internal representation of the highly nonlinear mapping from input to output. It could either be a classification problem where the correct label needs to be identified for a given input, or a regression problem where correct estimation of a response is desired.  An autoassociative network (autoassociator) is an artificial neural network formulation which tries to learn the reconstruction of input using backpropagation. Thus, for an autoassociator, the input is same as the output and an approximation to the identity mapping is obtained in a nonlinear setting. In the past, some researchers have used neural networks as autoassociators with the aim of extracting sparse internal representations of the input data (\emph{e.g.} \cite{Ackley1985, Cottrell1987, Chauvin1989}). However, the use of insufficient layers restricted the generalization of these networks, therefore limiting their applicability. \citet{Kramer1992} used three hidden layers comprising linear and nonlinear activations and showed the applicability of their autoassociator for gross noise reduction in process measurements. However, for cases where the features of a process are related through a complex nonlinear function, the use of even three hidden layers may not necessarily be sufficient for certain cases.

\citet{Bengio2009} presented \emph{autoencoder}, a form of autoassociators in a deep network framework, which allowed learning more accurate internal representations of the input. Since `autoencoder' is a more common term in the recent literature, the rest of the paper uses it over `autoassociators'. Autoencoders have primarily been used to reduce the dimensionality of large datasets. A projection to a lower dimensional space helps to identify several hidden features and promotes improved interpretation of the data. \citet{Vincent2010} used autoencoders for denoising tasks by cheaply generating input training data and corrupting it. This denoising procedure was aimed at making autoencoders more robust and allowed reducing the dimension of data efficiently even in the presence of noise. \citet{Valentine2012} presented the geophysical application of autoencoders for data reduction and quality assessment of waveform data. The paper presented a precise and clear overview of the classical autoencoder theory, followed by its application on seismic waveform data. Since then, autoencoders have already been used for a few other problems \emph{e.g.} analysis of topographic features \cite{Valentine2013}, identification of geochemical anomalies \cite{Xiong2016}.  

\indent Amongst several others, one of the biggest challenges in geophysics is the \emph{denoising} of data. The problem of noise removal from data is very common with various other disciplines, and has been studied extensively in the past. Some of these are based on local smoothing to blur the noise \emph{e.g.} nonlinear total variation based approaches \cite{Rudin1992}, anisotropic diffusion method \cite{Weickert1998},  bilateral filtering \cite{Tomasi1998}, \emph{etc}. Other category of denoising approaches involves learning on noise-free datasets and then exploitation of noisy datasets \cite{Weiss2007, Jain2008, Roth2009}. One such way is to learn using wavelets and then shrinkage of the coefficients to remove the noise \cite{Pizurica2002, Portilla2003}. Wavelet based shrinking has been used on various geophysical problems, \emph{e.g.} seismic noise attenuation \cite{Beenamol2012, Li2017}. 

In the context of denoising, limited research has been done in the past to investigate the applicability of autoencoders. Recently, \citet{Burger2012} used autoencoders in a deep network framework for denoising input images and the approach was found to outperform some and perform equal to other state-of-art denoising methods. \citet{Schuler2013} presented a neural network based non-blind image deconvolution approach capable of sharpening blurred images. The network was trained on large datasets comprising noisefree as well as noisy images, and it was found to work well for the task of denoising. \citet{Ojha2016} used autoencoders for denoising high-resolution multispectral images. It was shown in this paper that after training the model on a large set of noisy and denoised images, results comparable to non-local means algorithm 
are obtained in a significantly lesser amount of time. 

Clearly, the works on denoising outlined above have demonstrated the potential of autoencoders. In a recent work, we have briefly shown that denoising autoencoders work very well for geophysical problems \cite{Bhowmick2018}, and it is of interest to explore further in this direction. In this paper, we study in detail the application of autoencoders for denoising geophysical data. This work revolves around using autoencoders to learn the representation of the signal, and separate the noise content. We start with exploiting the potential of shallow autoencoders for denoising purpose. Based on the identified limitations of these networks, deep autoencoders with several different number of hidden layers are tested. To further enhance the denoising characteristic of the autoencoders, a stacked formulation is presented, the application potential of which is demonstrated on various numerical examples.

Here, we discuss the outline for the rest of the paper. The theoretical details of autoencoders are discussed in \mbox{Section \ref{sec_theory}}. This includes a brief description of the traditional autoencoders (Section \ref{sec_theorya}) followed by its denoising variant (Section \ref{sec_den_autoenc}). The concept behind the stacking of autoencoders is discussed in Section \ref{sec_theoryc}. To demonstrate the working of autoencoders for denoising tasks, a basic mathematical example is presented in Section \ref{motiv_ex}. The applications on geophysical problems are discussed in Section \ref{sec_appgeo} and the final discussions and conclusions are presented in Sections \ref{discuss} and \ref{conclude}, respectively.  
 
\section{Theory}
\label{sec_theory} 
\subsection{Autoencoder}
\label{sec_theorya}
\indent  Autoencoders aim at learning the internal representation of data, typically encoding, and identify the important hidden features \cite{Bengio2009}. In its simplest form, an autoencoder is very similar to a multilayer perceptron (MLP), which consists of an input layer, an output layer, and one or more hidden layers. For an ideal autoencoder, the input and output are same, which implies that the hidden units need to be tuned such that an accurate nonlinear approximation to the identity function can be obtained. 

\begin{figure}
    \centering
    \includegraphics{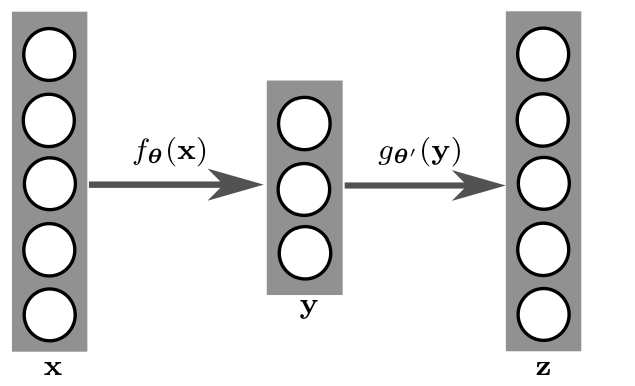}
    \caption{Schematic structure of a traditional neural network. For $\mathbf{x} \approx \mathbf{z}$, the network corresponds to an autoassociator/autoencoder.} 
    \label{fig_basic_ae}
\end{figure}

When neural networks are used, our interest is in learning an internal representation that relates the input to the output. Fig. \ref{fig_basic_ae} shows the schematic diagram of a neural network, where $\mathbf{x}$ and $\mathbf{z}$ are the input and output vectors, respectively. The output vector $z$ is obtained from $\mathbf{x}$ through a series of linear/nonlinear mappings denoted by $f_{\boldsymbol\theta}(\cdot)$ and $g_{\boldsymbol\theta'}(\cdot)$ functionals, respectively. For a deep network, these mappings themselves could comprise several hidden layers. The vector $\mathbf{y}$ in Fig. \ref{fig_basic_ae} corresponds to a hidden layer in the network with reduced dimensionality. 

For the network shown in Fig. \ref{fig_basic_ae} to be formulated as an autoencoder, $\mathbf{x}$ and $\mathbf{z}$ need to be ideally the same. Generally, the dimensionality of the hidden layers (\emph{e.g.} $\mathbf{y}$) is  kept lower than that of the input and output, which allows autoencoders to learn an approximate compressed representation of the input. As discussed in \cite{Vincent2010}, a natural criterion that any good representation should be expected to meet is that a significant amount of information about the input is retained. However, this condition alone is not sufficient to yield a good representation. Using hidden layers of same dimensionality or higher can lead to identity mapping, which is unlikely to lead to any useful information. Although traditionally followed, it is not necessary to use hidden layers of lower dimensions, rather, these can be larger than the input.  

With the constraint of reduced dimensionality in the hidden layers, autoencoders can provide an alternative reduced-dimensional representation of the massive datasets, providing a novel insight into the data. Adding sparsity constraint allows using higher dimensionalities and it has been observed that these can provide very useful feature representations (\emph{e.g.} \cite{Ronzato2007}). An advantage of the sparse autoencoders is that they can handle variable-sized representations \cite{Vincent2010}. A simplified version of an autoencoder, where no nonlinear transformations are used and a squared-loss error function is employed, is equivalent to performing principal component analysis (PCA) \cite{Baldi1989}. However, this is generally not true for the traditional autoencoders where sigmoid-based nonlinearity exists. 

\indent A traditional autoassociator consists of two parts: an \emph{encoder} and a \emph{decoder}. Looking back at Fig. \ref{fig_basic_ae}, let us assume that the shown neural network corresponds to an autoencoder with one hidden layer. Thus, $\mathbf{y}$ corresponds to the hidden representation with reduced dimensionality. The mapping phase where the input $\mathbf{x}$ is transformed into the hidden representation $\mathbf{y}$ is termed as an encoder. A decoder is the part of an autoencoder where the input is reconstructed back as $\mathbf{z}$ from its hidden representation $\mathbf{y}$. In Fig. \ref{fig_basic_ae}, the encoding and decoding functions are denoted by $f_{\boldsymbol\theta}(\cdot)$ and $g_{\boldsymbol\theta'}(\cdot)$, respectively and these mappings are parametrized by vectors $\boldsymbol\theta$ and $\boldsymbol\theta'$, respectively. Typically, the mapping functions comprise of affine mapping followed by certain nonlinearity and can be expressed as:
\begin{align}
& f_{\boldsymbol\theta}(\mathbf{x}) = \mathcal{S}(\mathbf{Wx + b}), \\
& g_{\boldsymbol\theta'}(\mathbf{y}) = \mathcal{S}(\mathbf{W'y + b'}),
\end{align}
where, $\boldsymbol\theta = \{\mathbf{W, b}\}$ and $\boldsymbol\theta' = \{\mathbf{W', b'}\}$ are parameter sets with $\mathbf{W}$ and $\mathbf{W'}$ denoting the weight matrices and $\mathbf{b}$ and $\mathbf{b'}$ representing the bias vectors, respectively. Typically the nonlinear mapping $\mathcal{S}(\cdot)$ is achieved using sigmoid or radial basis functions.

\indent The goal of the autoencoder presented in Fig. \ref{fig_basic_ae} is to minimize the reconstruction loss between $\mathbf{x}$ and $\mathbf{z}$. As error (loss) function, typically squared-error function or cross-entropy  loss are used depending on the type of problem. The autoencoder presented in Fig. \ref{fig_basic_ae} consists of a single hidden layer. However, for autoencoders of higher complexity, several hidden layers can be used. Accordingly, the encoder and decoder will then comprise of a series of mappings in each. For more details related to autoencoders, see \cite{Vincent2010}. 

\subsection{Denoising autoencoder}
\label{sec_den_autoenc}

\begin{figure*}
    \centering
    \includegraphics{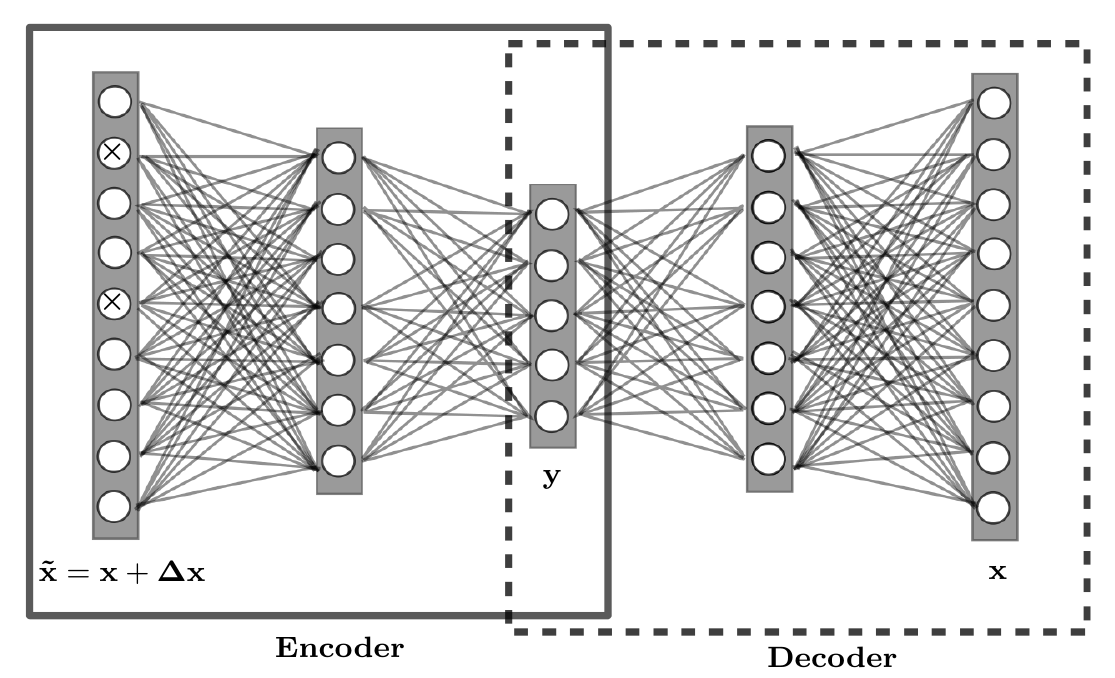}
    \caption{Schematic diagram of a denoising autoencoder showing the encoder and decoder segments. The network comprises 3 hidden layers of 7, 5 and 7 neurons, respectively.} 
    \label{fig_deauto}
\end{figure*}

Any model can be considered as a good denoiser, if it can output clean signal from a noisy input. The first and foremost thing needed to build a denoising autoencoder is to identify a mapping from the noisy domain to the noisefree domain. The complexity of this mapping depends on several factors (\emph{e.g.} level of noise), and cannot be expressed using a simple formula. However, if there exist a large number of data samples, autoencoders can be used to determine the fitting empirical model \cite{Burger2012}.

As outlined in \cite{Vincent2010}, the concept of denoising autoencoders is based on the following ideas:
\begin{itemize}
\item The higher level representation of the input data (\emph{i.e.} primary signal) is generally assumed to be more stable and robust to addition of noise.
\item Denoising approach should be able to capture the features associated with the primary signal in the inner hidden layers of the network.
\end{itemize}
Typically, it is assumed that our primary signal has a certain well-defined representation, while noise may have it or may not. For non-coherent noise, the denoiser needs to be trained to filter our the components of input data which do not comprise any well-defined pattern. For coherent noise, the denoiser has to be trained such that it preserves the representation of the signal, but filters out the noise component. Removing coherent noise can be a challenging problem, especially when the signal-to-noise ratio is quite low.

Fig. \ref{fig_deauto} shows the schematic network representation for a basic denoising autoencoder. The autoencoder comprises three hidden layers of 7, 5 and 7 neurons respectively, and the input consists of 9 features. Compared to the input,the dimensionality of the innermost representation is 44\% lower, which means that a compressed representation of the input will be encoded, and there can possibly be a loss of certain features. The goal is to train this network to construct clean output $\mathbf{x}$ from the corrupted version $\mathbf{\tilde{x}}$. 

Through a series of two projections (let us assume $f_{\boldsymbol\theta}(\mathbf{\tilde{x}})$), the noisy signal $\mathbf{\tilde{x}}$ is mapped onto a reduced dimensional space, and the result is $\mathbf{y}$. The functional $f_{\boldsymbol\theta}(\cdot)$ here refers to the encoder part of the autoencoder. Ideally, from the hidden representation $\mathbf{y}$, the noisefree signal $\mathbf{x}$ needs to be constructed, and this process is referred to as decoding ($\mathbf{x} = g_{\boldsymbol\theta'}(\mathbf{y})$). During the optimization process, this is achieved by training 
the set of parameters $\boldsymbol\theta$ and $\boldsymbol\theta'$ and obtaining output $\mathbf{z}$, such that the reconstruction error $\epsilon(\mathbf{x, z})$ is minimized. 

\citet{Vincent2010} have provided a nice geometrical interpretation for denoising autoencoders. This interpretation is based on the so-called \emph{manifold assumption} (\cite{Chapelle2006}), according to which natural high-dimensional  data concentrates close to a non-linear low-dimensional manifold. Generally, the noisefree data can be understood as a combination of several principal components, and using a neural network architecture, it is possible to obtain these components. Noise is generally found to be shifted away from these manifolds, and in the process of minimizing the loss, the optimization process tends to not include the noise component in the reduced dimensionality. 

\subsection{Stacked autoencoders}
\label{sec_theoryc}

\begin{figure*}
    \centering
    \includegraphics{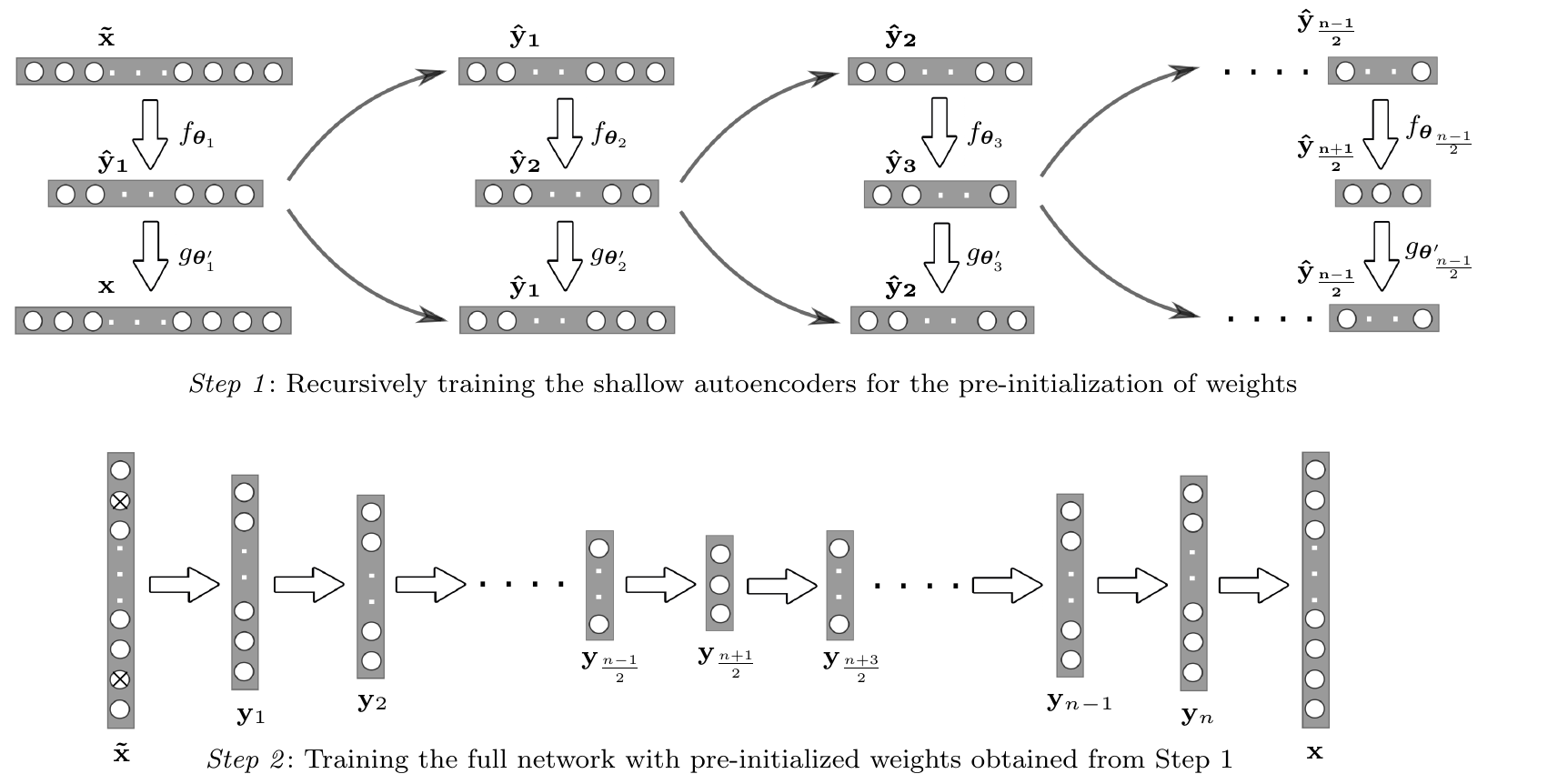}
    \caption{Schematic diagram of a stacked denoising autoencoder showing the two steps involved in the denoising process. The entire deep network is assumed to comprise $n$ hidden layers. During Step 1, the weights are trained for one hidden layer at a time, in a recursive manner. Finally, during Step 2, the entire network is trained at once using the pre-initialized weights obtained from Step 1.} 
    \label{fig_stack_ae}
\end{figure*}

Training the autoencoder to remove noise from a given dataset can be highly nonlinear and is not an easy problem to solve. Learning such complex representations requires deep multi-layered neural networks. The standard approach, comprising random initialization of weights and using gradient descent based backpropagation, is known to produce poor solutions for 3 or more hidden layers. In \cite{Larochelle2007}, this aspect has been studied in detail, and it has been observed that if efficient algorithms are used, deep architectures perform better than shallow ones.	

In this paper, a stacked formulation for autoencoders is proposed, where multiple cycles of simple autoencoders are trained in a zoom-in fashion, followed by training the whole network at once. Fig. \ref{fig_stack_ae} shows the schematic diagram explaining the two steps of stacked denoising autoencoder. It is assumed that the deep network architecture of the autoencoder comprises $n$ hidden layers. The details related to the two steps follow below.
\subsubsection{Recursive pre-training of weights and biases}
In this step, the weights and biases of the network are pretrained using simple autoencoders comprising 1 hidden layer each. For $n$ hidden layers in the deep autoencoder network, a total of $(n+1)/2$ autoencoders need to be formulated due to symmetry in the network architecture. Here, it is assumed that $n$ is an odd number as shown in Fig. \ref{fig_stack_ae}. The weights corresponding to the hidden layers are determined, starting with the outermost hidden layers and moving towards the innermost representation in a recursive manner. 

\indent Let $\mathbf{\tilde{x}}$ and $\mathbf{x}$ denote the noisy signal and its uncorrupted version, respectively. Let $f_{\boldsymbol\theta_1}(\cdot)$ and $g_{\boldsymbol\theta'_1}(\cdot)$ denote the projection of $\mathbf{x}$ onto the first hidden layer space (where the representation is denoted by $\mathbf{\hat{y}}_1$) and projection from the hidden layer to the output space, respectively. This implies that $\mathbf{\hat{y}}_1 = f_{\boldsymbol\theta_1}(\mathbf{\tilde{x}})$ and \mbox{$\mathbf{x} = g_{\boldsymbol\theta'_1}(\mathbf{\hat{y}}_1)$}. Note here that after the autoencoder has been trained, there will still be an approximation error, and $ g_{\boldsymbol\theta'}(\mathbf{\hat{y}}_1)$ may not necessarily be equal to $\mathbf{x}$. Once the parametrization vectors $\boldsymbol\theta_1$ and $\boldsymbol\theta_1'$ have been trained, the next inner representation needs to be optimized.

For training the next level representation, an autoencoder comprising a 3-layer neural network is formulated. The input and output vectors for this autoencoder are set to $\mathbf{\hat{y}}_1$ each. The respective mappings onto the hidden space and the output space are denoted by $f_{\mathbf{\boldsymbol\theta}_2}(\cdot)$ and $g_{\mathbf{\boldsymbol\theta'}_2}(\cdot)$ functionals, respectively. The parameter vectors $\boldsymbol\theta_2$ and $\boldsymbol\theta_2'$ are optimized, and $\mathbf{\hat{y}}_3$ is computed. This whole process of formulating an autoencoder and computing level hidden level representation is repeated $\frac{n+1}{2}$ times as shown in Fig. \ref{fig_stack_ae} and $\mathbf{\hat{y}}_1, \mathbf{\hat{y}}_2, \hdots, \mathbf{\hat{y}}_{\frac{n+1}{2}}$ are obtained. At the same time, parameter vectors $\boldsymbol\theta_1, \boldsymbol\theta_2, \hdots, \boldsymbol\theta_{\frac{n+1}{2}}$ and  $\boldsymbol\theta_{\frac{n+1}{2}}', \boldsymbol\theta_{\frac{n-1}{2}}', \hdots, \boldsymbol\theta_1'$ have been optimized to certain values.

\subsection{Training the full network}
Once the parametrization vectors have been initialized, Step 2 involves further training the entire deep network at once. The deep network architecture has been shown in Fig. \ref{fig_stack_ae}. The input and output vectors for this network are set to $\mathbf{\tilde{x}}$ and $\mathbf{x}$, respectively and the parameters for the layers from left to right are set to $\boldsymbol\theta_1, \boldsymbol\theta_2, \hdots, \boldsymbol\theta_{\frac{n+1}{2}}, \boldsymbol\theta_{\frac{n+1}{2}}', \boldsymbol\theta_{\frac{n-1}{2}}', \hdots, \boldsymbol\theta_1'$.  Accordingly, the mapping functionals from left to right of the network are defined to be $f_{\mathbf{\boldsymbol\theta}_1}(\cdot)$, $f_{\mathbf{\boldsymbol\theta}_2}(\cdot)$, $\hdots$, $f_{\mathbf{\boldsymbol\theta}_\frac{n+1}{2}}(\cdot)$, $g_{\mathbf{\boldsymbol\theta}'_\frac{n+1}{2}}(\cdot)$, $g_{\mathbf{\boldsymbol\theta}'_\frac{n-1}{2}}(\cdot)$, $\hdots$, $g_{\mathbf{\boldsymbol\theta}'_1}(\cdot)$, respectively. Once the whole network has been set up and the parameters have been initialized with the values computed in Step 1, optimization is performed and the new representations for the $n$ hidden layers of the deep network are obtained. These are denoted by $\mathbf{y_1}, \mathbf{y_2}, \cdots, \mathbf{y_n}$.

\section{A basic mathematical example}
\label{motiv_ex}

To provide a better understanding of how an autoencoder works, here we present a basic mathematical example. Since the goal of this paper is typically to demonstrate the use of autoencoders for noise reduction and data reconstruction and excludes its usage as merely a dimensionality reduction tool, we restrict ourselves to denoising autoencoders. For readers interested in the application of autoencoders for dimensionality reduction in geophysics, we advise looking at the work of \cite{Valentine2012}.

To start with, a simple mathematical problem of two parameters ($z$ and $\theta$) is chosen and a process model of the following form is used \cite{Gupta2012},
\begin{align}
& x = tz\sin\theta + \frac{t^2}{z}\cos\theta, \qquad t = 0.05:0.05:1, \nonumber \\
& z \in [0.5, 4.0], \qquad \theta \in [0.3, 1.3] \textrm{ in radians.},
\label{eq_proc1}
\end{align}
where, $t$ and $x$ are the input and output fields, respectively, and $z$ and $\theta$ are the model parameters. For $t$, a range of 0.05 to 1 is chosen with a sampling interval of 0.05. The input field $\mathbf{t} = \{0.05, 0.1, \hdots, 1.0\}$ is then used to generate the output signal $\mathbf{x} = \{x_1, x_2, \hdots, x_n\}$, where $n$ denotes the number of sampling points. Once $\mathbf{x}$ is obtained, it is assumed that the process model is not known anymore. Next, an autoencoder is trained to learn the internal representation of $\mathbf{x}$ such that for any noisy variant $\tilde{\mathbf{x}}$, the noise-free signal can be recovered. Learning the representation here refers to approximating the process model through a neural network and rejecting the component of the input which does not fit well with the model. 

An autoencoder comprising 2 hidden layers is formulated. The number of neurons in each hidden layers is kept to be equal to $n$. Note that choosing the number of neurons equal to the number of input units here does not lead to a plain identity function due to the large noise added in some of the signals.  Table \ref{motiv_ex} states the neural network parameters used for this autoencoder. Nonlinear (sigmodial) activation functions are used for projection from the input layer to hidden layer 1 and hidden layer 1 to hidden layer 2. For obtaining the output $\mathbf{z}$, linear  activations are used. The error (loss) function $\mathcal{J}(\cdot, \cdot)$ is defined as 
\begin{equation}
\mathcal{J}(\mathbf{x}, \mathbf{z}) = \sum_{i=1}^{N_s}\frac{(\mathbf{x - z})^{\intercal}(\mathbf{x - z})}{N_s},
\label{eq_loss_1}
\end{equation} 
where, $N_s$ refers to the number of samples,and  $\mathbf{x}$ and $\mathbf{z}$ refer to the noise-free and recovered samples, respectively.

A set of 20000 samples is generated using the process model stated in Eq. \ref{eq_proc1}, and is further divided into 80\% and 20\% for training and validation samples, respectively. Random noise of upto 25\% is added to the data points of 10000 samples and the other 10000 samples are kept noise-free. For 50\% of the noisy-samples, the magnitude of added noise scales based on the local value at the respective point of the sample. For the remaining 50\%, it scales with the mean of all the data points of the sample. For optimization purpose, the traditional gradient descent algorithm is used. Due to the simplicity of the problem, no regularization is needed, and the convergence of the optimization problem is found to be very fast.

\begin{table}
\caption{Parameters for the simple denoising autoassociative neural network used in Section \ref{motiv_ex}.}
\centering
\begin{tabular}{ l | c }
  Parameter & Value \\
  \hline
  TRAINING PHASE & \\
  No. of samples & 20000 \\
  Noisy samples & 10000 \\
  Noise type & random noise (upto 25\%) \\
  Training samples & 80\% \\
  Validation samples & 20\% \\
  No. of hidden layers & 2 \\
  Hidden units & \{20, 20\} \\
  Activation functions & \{sigmoid, sigmoid, linear\} \\
  \hline
  TEST PHASE & \\
  Test samples & 100 \\
  noise & between 10\% and 25 \% \\
  \hline

\end{tabular}
\end{table}
\begin{figure*}
\centering
	\begin{subfigure}{0.32\linewidth}
	\begin{center}
	\includegraphics{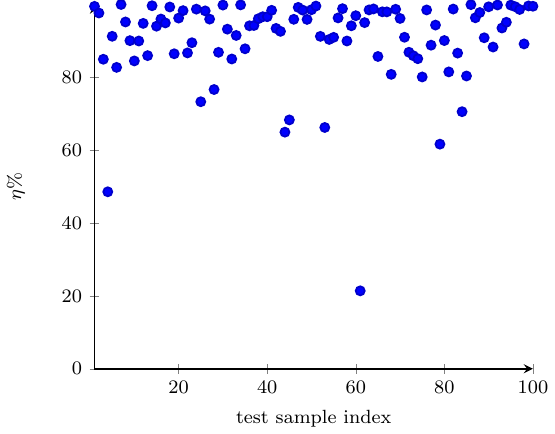}
    \caption{Noise reduction $\eta$ for the test data}
	\label{fig_eta_plot}
	\end{center}
	\end{subfigure}
	\begin{subfigure}{0.32\linewidth}
	\begin{center}
	\includegraphics{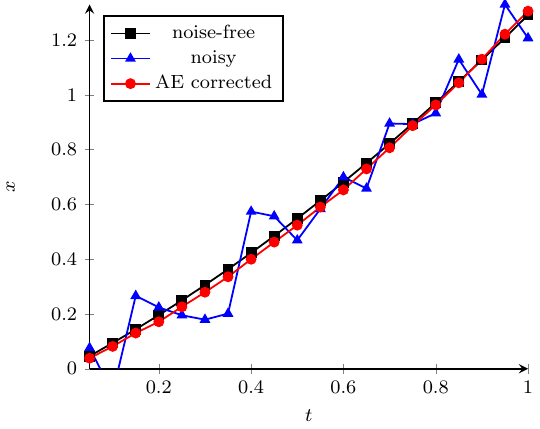}
    \caption{Test sample 20}
    	\label{fig_plot20}
	\end{center}
	\end{subfigure}
	\begin{subfigure}{0.32\linewidth}
	\begin{center}
	\includegraphics{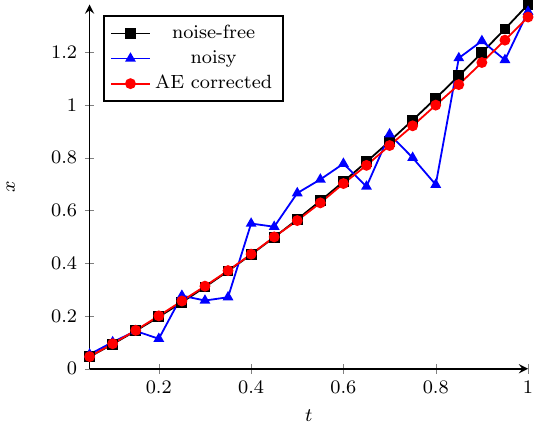}
    \caption{Test sample 70}
    	\label{fig_plot70}
	\end{center}
	\end{subfigure}
	\caption{(a) Relative noise reduction $\eta$ for the 100 noisy test samples, (b) noise-free, noisy and autoencoder (AE) corrected data for sample index 20 and (c) for sample index 70. A denoising autoencoder  comprising 2 hidden layers with 20 neurons in each is used.}
\end{figure*}
\indent To test the accuracy of the learnt representation $\mathbf{\Phi}$, 100 test samples are generated, and noise, chosen randomly between the levels of 10\% and 25\%, is added to each of these samples. Next, these data samples $\mathbf{\tilde{x}}$ are passed through the learnt representation to reduce the noise and obtain the output $\mathbf{z = \Phi(\tilde{x})}$. The efficiency $\eta$ of the learnt representation $\mathbf{\Phi}$ for any noisy signal $\mathbf{\tilde{x}}$ is given by 
\begin{equation}
\eta  = 100 \times \frac{(\mathbf{z - x})^{\intercal}(\mathbf{z - x})}{(\mathbf{\tilde{x} - x})^{\intercal}(\mathbf{\tilde{x} - x})} \enskip \%.
\label{eq_cost_fn}
\end{equation}

Fig. \ref{fig_eta_plot} shows the $\eta$ values for 100 noisy samples measured using Eq. \ref{eq_cost_fn}. For these samples, the mean value of $\eta$ is found to be $\eta = 90.27\%$, which means that the learnt autoencoder $\mathbf{\Phi}$ can reduce the noise by approximately 90\%. This is a significant improvement and clearly demonstrates the applicability of autoencoders for denoising purpose. Figs. \ref{fig_plot20} and \ref{fig_plot70} show two data samples, their noisy versions as well as the autoencoder corrected signals. These data samples have been picked randomly from the set of 100 samples for demonstration purpose	. From the results, it is clear that the autoencoder has learnt to identify the signal pattern within the data, and reject any random noise.

\section{Applications}
\label{sec_appgeo}
\subsection{Self-potential problem}

\begin{figure}
    \centering
    \includegraphics{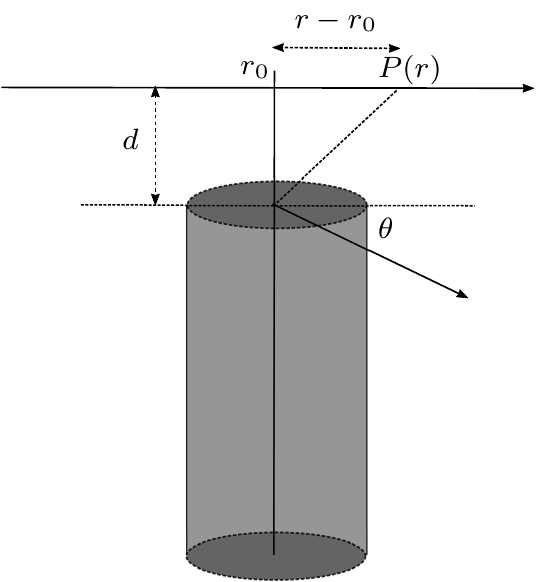}
    \caption{Schematic diagram of a buried vertical cylinder, also showing some of the parameters that characterize the 1D SP anomaly caused due to it.} 
    \label{fig_sp_cylinder}
\end{figure}

In a self-potential (SP) survey, the naturally occurring potential differences generated by electrochemical, electrokinetic and thermoelectric sources are measured. This approach has been used in a wide range of applications: in exploration mainly for sulphides and graphite \cite{Sundararajan1998}, ground water investigations \cite{Santos2002}, detection of cavities   \cite{Jardani2007} and geothermal exploration \cite{Zlotnicki2003}. For some cases, SP anomaly can be modeled using simple geometries, \emph{e.g.} sphere, cylinder and inclined sheet \cite{Santos2010}. From the obtained field data, the set of parameters defining the buried source can be determined using various methods such as curve matching \cite{Meiser1962}, gradient-based methods \cite{Abdelrehman2003}, global optimization \cite{Gupta2013}, \emph{etc.}

The SP data acquired in the field can also comprise noise from several sources, and without any post-processing, it is possible that the inverted set of parameters defining the buried source do not comply well with the actual properties. With the gradient-based methods, there exist high chances of getting stuck in a local optimum. The global optimization methods as well have been found to be sensitive to the level of noise in the data (\emph{e.g.} \cite{Santos2010, Gupta2013}). To circumvent this problem, we use autoencoders for reducing the noise content in the acquired data. 

In the context of SP anomaly inversion, the application of autoencoders is demonstrated on data obtained for a 1D problem. A vertical cylinder buried in the subsurface, as shown in Fig. \ref{fig_sp_cylinder}, is considered. The forward model for the SP anomaly at a point $P(r)$ is computed as
\begin{equation}
V(r) = K \frac{(r - r_0)\cos\theta + d\sin\theta}{((r - r_0)^2 + d^2)^q}, 
\label{eq_fw_sp}
\end{equation}
where, $d$, $\theta$, $K$ and $q$ denote depth, polarization angle, current dipole moment and shape factor, respectively, and $r_0$ refers to the origin of the anomaly. These five variables are the parameters obtained generally by inverting the SP data. Note that in this work, we are not developing another efficient inversion approach. Rather, with the forward model known, our goal is to train the neural network to be able to identify the component of data that complies with it, and reject the other parts. 

\begin{table}
\caption{Range of values for the parameters characterizing the forward model for 1D SP anomaly caused due to the burial of a vertical cylinder (\cite{Santos2010}).}
\centering
\begin{tabular}{ l | c | c }
  Parameter & Min  & Max\\
  \hline
  depth $d$ (m) & 1 & 8 \\
  polarization angle $\theta$ (degrees) & 25 & 75 \\
  electric dipole moment $K$ (mV) & -1000 & 1000 \\
  shape factor $q$ & 0.5 & 1.5 \\
  origin of the anomaly $r_0$ (m) & -5 & 5 \\
  \hline
\end{tabular}
\label{table_sp_ranges}
\end{table}

To start with, we define ranges for the parameters stated in Eq. \ref{eq_fw_sp}, and these are shown in Table \ref{table_sp_ranges}. Combinations of parameters' values are randomly chosen from these ranges to generate data samples for training the neural network. The value of $r$ is varied from -20.0 m to 20.0 m with a spacing of 2.5 m. As stated in Table \ref{table_sp_nn1}, a total of 60000 samples are used, out of which 40000 samples are corrupted with random noise. The entire dataset is divided into training and test sets in the ratio of 4:1. All the layers comprise sigmoid activations, except the last one which has only a linear activation. The loss (error) at every step of training is computed in a similar fashion as stated in Eq \ref{eq_loss_1}. Further, to get the quantitative estimate of the noise reductions, we use the function $\eta$ stated in Eq. \ref{eq_cost_fn}.
\begin{table}
\caption{Network parameters for the various (stacked) denoising autoencoders used for reduction of noise in self-potential data.}
\centering
\begin{tabular}{ l | c }
  Parameter & Value \\
  \hline
  TRAINING PHASE & \\
  No. of samples & 60000 \\
  Noisy samples & 40000 \\
  Noise type & uniform random noise \\
  Noisy units per sample & 50\% \\
  Training samples & 80\% \\
  Validation samples & 20\% \\
  Activations & all sigmoid and last as linear\\
  \hline
  TEST PHASE & \\
  Test samples & 1000 \\
  Noisy units per sample & 50\% \\
  Noise & random (upto 50) \% \\
  \hline
\end{tabular}
\label{table_sp_nn1}
\end{table}

\begin{table*}
\caption{Information related to runs of denoising SP data using several autoencoder configurations. Here, SA, DA, SDA and SDA-R refer to shallow autoencoders, deep autoencoders, stacked deep autoencoders and stacked deep autoencoders with randomness, respectively. Also $\lambda$ defines the extent of regularization $\eta$ denotes percentage reduction in noise.}
\centering
\begin{tabular}{ c | c | c | c}
Hidden nodes & Network-type & $\lambda$ & $\eta$ (in \%)\\
  \hline
\{4\} & SA  & 0.0 & 17.6 \\
\{12\} & SA & 0.0 & 54.7 \\
\{25\} & SA & 0.0 & 55.4 \\
\{12, 12\} & DA & 0.0 & 58.4 \\
\{20 25, 20\} & DA & 0.0 & 65.4 \\
\{17, 20 25, 20\} & DA & 0.0 & 64.1 \\
\{20 25, 30, 25, 20\} & DA & 0.0 & 63.7 \\
\{20 25, 20\} & SDA & \{0.0, 0.0; 0.0\} & 61.8 \\
\{20 25, 20\} & SDA & \{0.05, 0.05; 0.0\} & 70.9 \\
\{20, 25, 30, 25, 20\} & SDA & \{0.05, 0.05; 0.05; 0.0\} & 73.0 \\
\{20, 25, 30, 25, 20\} & SDA-R & \{0.05, 0.05; 0.05; 0.0\} & 78.5 \\
\{25 20, 17, 20, 25\} & SDA & \{0.0, 0.0; 0.0; 0.0\} & 72.1 \\
\{35 25, 17, 25, 35\} & SDA & \{0.05, 0.05; 0.05; 0.0\} & 77.2 \\
\{35 25, 17, 25, 35\} & SDA-R & \{0.0, 0.0; 0.0; 0.0\} & 80.6 \\
\{35 25, 17, 25, 35\} & SDA-R & \{0.05, 0.05; 0.05; 0.0\} & 81.3 \\
  \hline
\end{tabular}
\label{table_sp_autos}
\end{table*}

Several different autoencoder configurations are tested to understand how complexity and composition of the neural networks affect the performance of autoencoders. Table \ref{table_sp_autos} lists the number of neurons in each layer for the various autoencoders used. It is observed that with shallow autoencoders (SA), which comprise up to two hidden layers, the reduction in noise is less than 60\%. For the network with one hidden layer comprising only 4 neurons, noise reduction level is merely around 17\%. This happens because such a compressed internal representation might not be enough to fully capture the signal pattern. Increasing the number of hidden layers to 12 already pushes the efficiency $\eta$ of the autoencoder beyond 50\%.   

The use of deep autoencoders (DE) with 3 or more layers has been observed to further improve the performance. With 3-5 hidden layers, $\eta$ reaches close to 65\% (Table \ref{table_sp_autos}). An interesting observation is that with increasing number of hidden layers, the value of $\eta$ reduces. This is because as the network grows, training it becomes more difficult due to the increased number of variables and vanishing gradients. Clearly, with these bottlenecks, the conventional deep networks are not the right solution to obtain very efficient autoencoders for the SP problem.

To circumvent the issue related to training the deep networks in a standard manner, we explore the application of stacked autoencoders for denoising SP data. Several neural network configurations are tested using the two step approach described in Fig. \ref{fig_stack_ae}. In the first step, the weights corresponding to every hidden layer are trained using basic autoencoders with only one hidden layer each. Once the weights have been initialized, a full-fledged deep neural network is trained to reach the final solution. Stacked autoencoders have been found to push $\eta$ beyond 70\%. This can be improved further by regularizing the weights and avoiding over-fitting. With 5 hidden layers, the stacked autoencoder could achieve efficiencies close to 78\%. 

\begin{figure*}
\centering
	\begin{subfigure}{0.32\linewidth}
	\begin{center}
	\includegraphics{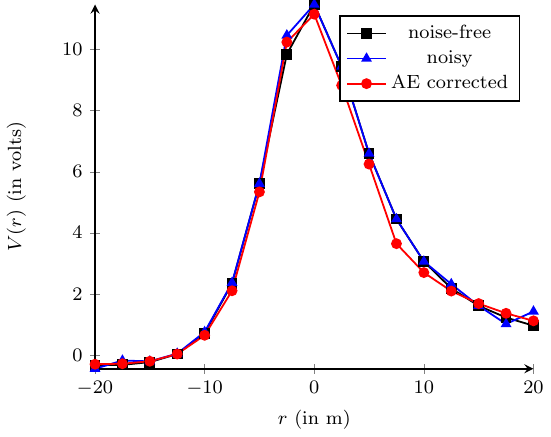}
    \caption{Test sample 27}
	\end{center}
	\end{subfigure}
	\begin{subfigure}{0.32\linewidth}
	\begin{center}
	\includegraphics{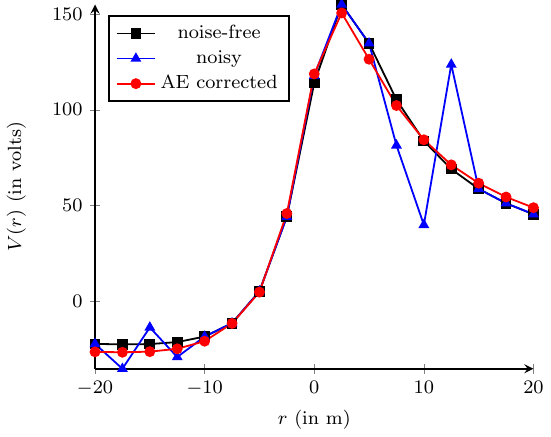}
    \caption{Test sample 73}
	\end{center}
	\end{subfigure}
	\begin{subfigure}{0.32\linewidth}
	\begin{center}
	\includegraphics{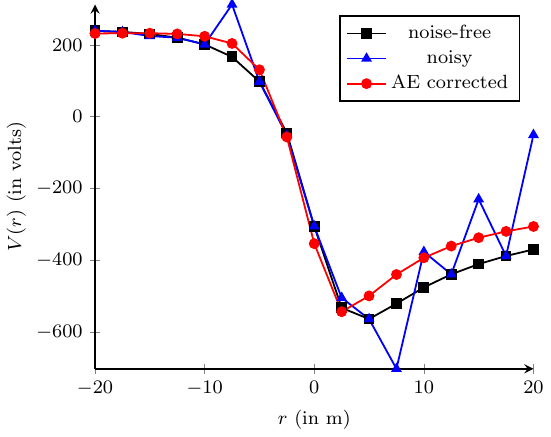}
    \caption{Test sample 282}
    \label{fig_sp_resultc}
	\end{center}
	\end{subfigure}
	\caption{Noise-free, noisy and autoencoder (AE) corrected data for samples with index 27, 73 and 282. These samples have been chosen randomly out of the 1000 test samples. The noisy signal consists of upto 50\% random noise in 50\% of the points for each sample. A stacked deep network with the hidden structure of \{25, 20, 17, 20, 25\} is used.}
	\label{fig_sp_plots}
\end{figure*}

Fig. \ref{fig_sp_plots} shows three data samples chosen randomly out of 1000 data samples in the test set. A stacked deep network with 5 hidden layers is used. It is observed that for the three cases, autoencoder could significantly reduce noise in the data. However, in Fig. \ref{fig_sp_resultc}, certain amount of bias can be seen in some parts of the result obtained using autoencoder. Although the employed autoencoder could smoothen the data in that region, the associated values deviate significantly from the actual values. A reason could be that the training set did not comprise samples resembling this data, and the model was not sufficiently trained for it. Clearly, a remedy would be to further train the model in a feedback loop based on the fitting obtained for such examples.   

We also observed that for stacked autoencoders, perturbing the weights obtained in step 1 improves the convergence of step 2. Randomly 10\% of the weights corresponding to each hidden layer are sampled and perturbed by up to 5\%. With this configuration, $\eta$ value of 80.6\% is obtained. This approach when combined with regularization can remove more than 81\% of the noise from SP data. With this level of improvement in the data, it can be claimed that stacked autoencoders could be a potential denoising tool for such problems.

\subsection{Seismic data}
\label{sec_seismic}
\begin{figure}
\centering
	\includegraphics{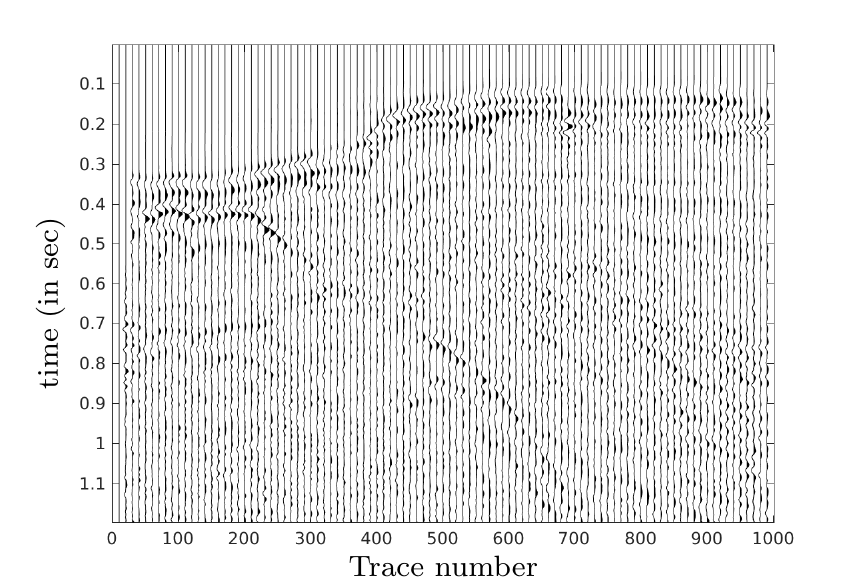}
\caption{An example seismic section considered for generating the training and test samples for this study.}
\label{fig_full_seismic}
\end{figure}

\begin{figure}
\centering
    \includegraphics{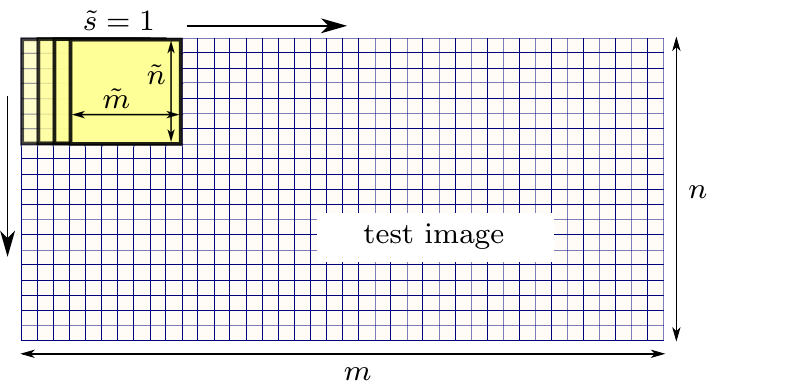}
\caption{Schematic diagram demonstrating the extraction of small image slices from a test image of size $m \times n$. For the seismic example case, $m$ and $n$ would denote time samples and number of traces, respectively. The image slices are chosen to be of size $\tilde{m} \times \tilde{n}$ and stride $\tilde{s} = 1$.}
\label{fig_image_stride}
\end{figure}
\begin{figure*}
\centering
	\begin{subfigure}{0.98\linewidth}
	\begin{center}
	\includegraphics{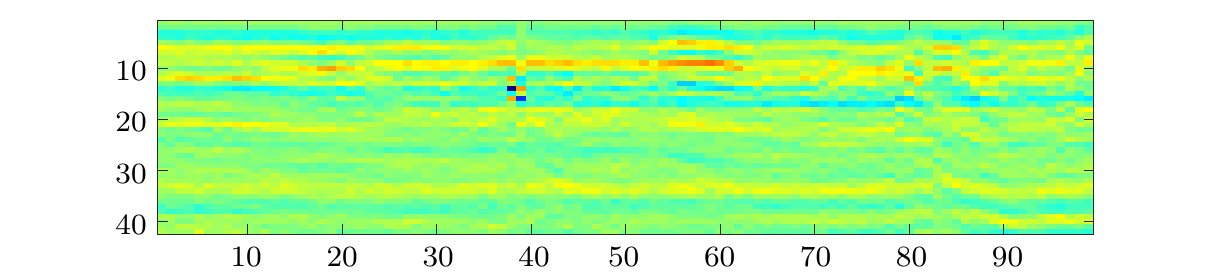}
    \caption{noise-free data}
    \label{fig_noisefree}
	\end{center}
	\end{subfigure}\\
	\begin{subfigure}{0.98\linewidth}
	\begin{center}
	\includegraphics{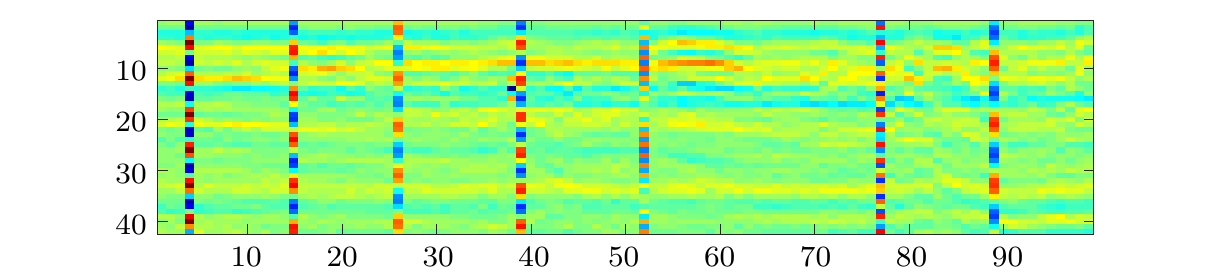}
    \caption{noisy data}
    \label{fig_noisy}
	\end{center}
	\end{subfigure}\\
	\begin{subfigure}{0.98\linewidth}
	\begin{center}
	\includegraphics{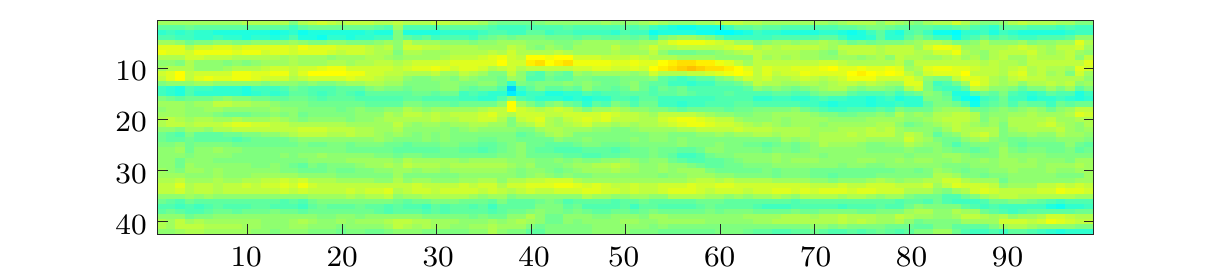}
    \caption{corrected data obtained using a traditional deep autoencoder}
    \label{fig_seismic_da}
	\end{center}
	\end{subfigure}\\
	\begin{subfigure}{0.98\linewidth}
	\begin{center}
	\includegraphics{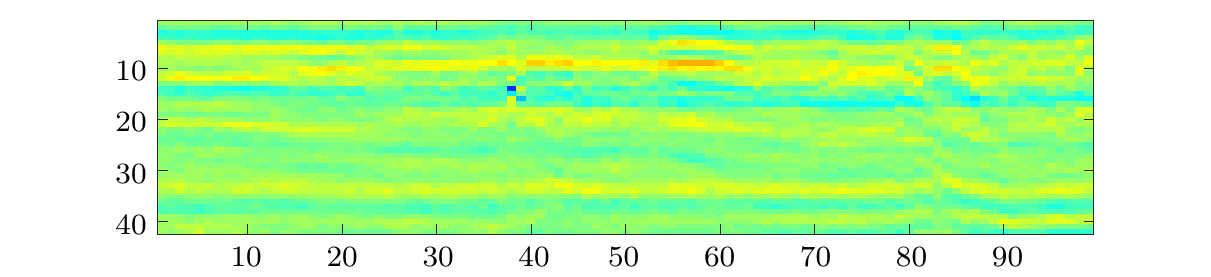}
    \caption{corrected data obtained using a stacked deep autoencoder}
    	\label{fig_seismic_sda}
	\end{center}
	\end{subfigure}
	\caption{Denoising of corrupted seismic data using a traditional deep autoencoder and a stacked deep autoencoder. For the noisy data, some traces chosen randomly, have been corrupted by replacing them with monofrequency sinusoidal traces in a frequency range of 100 to 220 Hz.}
	\label{fig_seismic_plots1}
\end{figure*}

In this section, the applicability of autoencoders is explored for the removal of random noise from seismic data. As stated earlier, the application of autoencoders on seismic waveform data has been demonstrated in the past by \cite{Valentine2012}, however, only in the context of dimensionality reduction. Here, our goal is to remove noise from seismic data. For simplification purpose, we do not discuss the headers associated with the data, rather we treat the seismic data only as a two-dimensional matrix of amplitude values. Also, in this paper, our scope is restricted to non-coherent noise, and cases of coherent noise are not considered. Fig. \ref{fig_full_seismic} shows the seismic section that has been used in this study to generate training and test samples for the autoencoder network.

\begin{table}
\caption{Network parameters for the various (stacked) denoising autoencoders used for reduction of noise in self-potential data.}
\centering
\begin{tabular}{ l | c }
  Parameter & Value \\
  \hline
  TRAINING PHASE & \\
  Sample size & $9 \times 42$ \\
  No. of samples & 1.1 million \\
  Noisy samples & 0.75 million \\
  Noise type & monofrequency sinusoidal noise \\
  Noise frequency & between 100 and 220 Hz \\
  Noisy traces per sample & 1 \\
  Training samples & 95\% \\
  Validation samples & 5\% \\
  Activations & all sigmoid and last as linear\\
  DA network & \{300, 400, 300\} \\
  SDA network & \{300, 400, 500, 400, 300\} \\
  Max. epochs & 50000 \\
  \hline
  TEST PHASE & \\
  Test image size & $99 \times 42$ \\
  Number of noisy traces & 7 \\
  Noise type & monofrequency sinusoidal noise \\
  Noise frequency & between 100 and 220 Hz \\
  window size & $9 \times 42$ \\
  stride & 1 \\

  \hline
\end{tabular}
\label{table_seismic_par1}
\end{table}

For the purpose of training the autoencoder, a total of 1.1 million small seismic samples are used. Details related to the training and test datasets are presented in Table \ref{table_seismic_par1}. From the seismic section shown in Fig. \ref{fig_full_seismic}, around 0.38 million smaller sample images are randomly chosen. Each sample image contains $42 \times 9$ data points, 9 being the number of traces and 42 denoting the number of data points along the time axis for every trace. These images are assumed to be the clean versions of data. Further, each uncorrupted image is used to generate two noisy samples. A trace is randomly chosen from the clean image, and it is replaced by a monofrequency trace with frequency in the range 100-220 Hz. The amplitude of the noisy trace is randomly chosen between $0.5 A_{max}$ and $A_{max}$, where $A_{max}$ refers to the maximum amplitude observed in the seismic section. In this way, a total of around 1.1 million seismic samples are obtained.  

The entire dataset is divided into training and validation sets in the ratio 95:5. To train the autoencoder, first we start with 3 hidden layers comprising 300, 400 and 300 neurons, respectively. All the activations are set to sigmoid, except the last one where linear activations are employed. The weights are initialized randomly and the entire network is trained for up to 50000 epochs. 

The trained model is then tested on a seismic image comprising 99 traces with 42 data points in each. \mbox{Fig. \ref{fig_noisefree}} shows the clean image used for generating test data. Table \ref{table_seismic_par1} lists complete details associated with testing the model. The noisy test data is generated by corrupting 7 traces in the noisefree test image as shown in Fig. \ref{fig_noisy}. The noisy traces correspond to monofrequency sinusoidal signals with frequency randomly chosen from the range 100-220 Hz. 

To feed the test image to the trained model, compatible image slices need to be chosen. These image slices are chosen using a window operator, as shown in Fig. \ref{fig_image_stride}. This window is slid along the row and column directions of the image using a certain stride $\tilde{s}$. Here, $\tilde{s}$ refers to the jump made by the window per step. For a test image size of $m \times n$, a window size of $\tilde{m} \times \tilde{n}$ and stride $\tilde{s} = 1$, a total of $\frac{m-\tilde{m}+1}{\tilde{s}} \times \frac{n-\tilde{n}+1}{\tilde{s}}$ image slices are fed to the trained model as shown in Fig. \ref{fig_image_stride}. Thus, the test seismic image used in this study is represented using 91 image slices. 

The output of the trained model is then summed up using a weighted approach to obtain the output test image of the seismic section. The image slices are properly aligned and stacked using weights proportional to the number of times a data point has been mapped into an image slice. Fig. \ref{fig_seismic_da} shows the denoised version of the noisy seismic image obtained using the autoencoder with 3 hidden layers. It is seen that the chosen autoencoder network could significantly remove random noise from the data. However, the autoencoder regularizes the seismic image due to which resolution of the image is lost to a certain extent. Clearly, this is not desired, since the reduced resolution will lose information related to thin beds as well as other fine features present in the seismic section.

Further, to test whether the random noise in seismic data can be reduced without compromising too much with the resolution of data, the potential of stacked autoencoder network is explored. A network comprising 5 hidden layers with 300, 400, 500, 400 and 300 neurons, respectively, is employed. In the first step of autoencoding, the weights are initialized using several traditional autoencoders comprising single hidden layer each. During the second step, the whole network is trained using the pre-initialized weights. The trained autoencoder is then used to denoise the test data. In the hidden layer comprising 500 neurons, an additional sparsity constraint is added which ensures that not all the neurons of this layer are activated at the same time. 

Fig. \ref{fig_seismic_sda} shows the denoised seismic image obtained from the trained stacked autoencoder.  The random noise has been significantly suppressed. Compared to Fig. \ref{fig_seismic_da}, it can also be seen that the resolution of the output image has significantly increased. However, with this autoencoder configuration as well, the resolution of output seismic data is compromised to a certain extent. Nevertheless, the stacked autoencoder configuration shows potential in suppressing noise in seismic data, and a future direction of research would be to design networks of even higher complexity that can produce better results.  

\begin{table*}
\caption{Range of values for the three property logs used to generate training dataset.}
\centering
\begin{tabular}{c | c | c | c }
Data source & Porosity (\%) & Clay fraction (\%) & Hydrate saturation (\%) \\
\hline
Synthetic 1  & 30 - 60 & 50 - 80 & 0 - 20 \\
KG basin (NGHP-01-05) & 0 - 90 & 85 - 95 & 0 - 30\\
Mt. Elbert-01, Alaska North Slope & 40 (approx.) & 0 - 40 & 0 - 60 \\
\hline
\end{tabular}
\label{well_table1}
\end{table*}

\subsection{Well log data}
\label{sec_well}

\begin{figure}
\centering
	\includegraphics{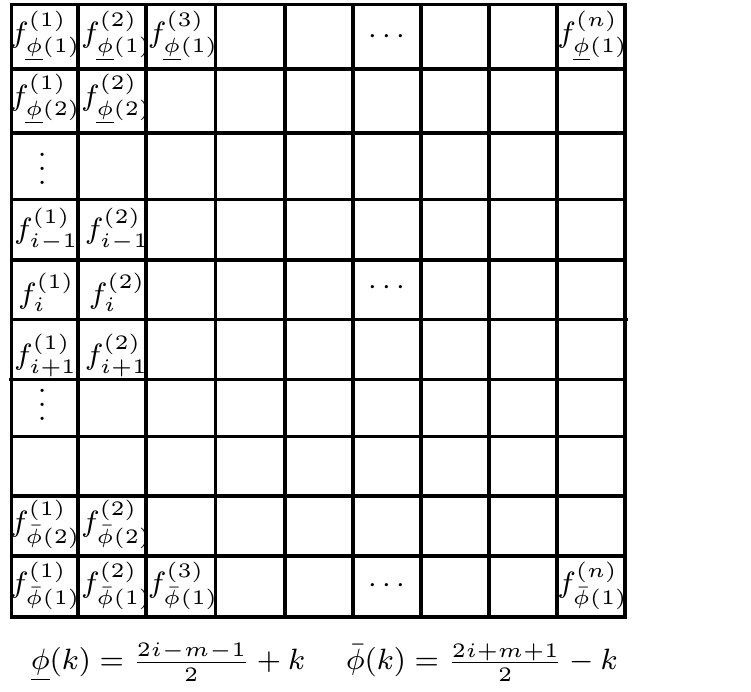}
\caption{Schematic diagram of an $m \times n$ image slice provided as a training sample to the denoising autoencoder. Here, the superscripts 1, 2, 3, $\hdots$, $n$ denote the features used for training. For the well data denoising problem considered in this paper, the features are porosity, saturation, p-wave velocity and clay content. To denoise the $i^{\text{th}}$ data point in the well logs, the image slice includes information of $\frac{m-1}{2}$ points above as well as below this data point in the well logs.}
\label{fig_well_train_image}
\end{figure}

Well logging is the practice of obtaining detailed information related to the geological formations in an area through sensors deployed in a borehole. This approach has widely been used to search for oil and gas, ground water, minerals as well as geotechnical studies. Several different properties such as porosity, density, velocity, water saturation, \emph{etc.} can be estimated using well log data. Often for a certain geology, empirical linear/nonlinear relationships are established between two or more such properties, and this is further used as a template to calculate one property if the other is known. A detailed discussion on some such templates can be found in \cite{Mavko2009} and references therein.

Amongst a set of logs acquired in a borehole, it is possible that information in some parts of one log is either corrupted or missing. In case the data is missing, it can be calculated from other logs using appropriate empirical relationships. Here, the challenge is to identify the correct empirical relationship, since it depends very much on the local geology. For cases where one of the logs contains some random noise, it might not be easy to identify it. 

In this paper, we explore the application of denoising autoencoders to solve the two issues outlined above. For the first case, we look at a suite of logs where parts of the data has been corrupted. Next, we train an autoencoder and use it to correct the data contained within this suite. In another test, parts of one of the logs are muted, and we let the trained autoencoder predict information in those parts. For study purpose, we use well log data from two gas hydrate sites: NGHP-01-05 site in the Krishna-Godavari (KG) basin of India \cite{Shankar2013} and Mt. Elbert (ME) site in the Alaska North Slope region \cite{Rose2011}.

For the two sites, the available logs are porosity ($\phi$), shale volume ($V_{\text{sh}}$) and $p$-wave velocity ($V_p$). Using these logs, gas hydrate saturation values ($S_h$) are calculated using the velocity-porosity transform proposed in \cite{Raymer1980}. It comprises 3 relations between velocity and porosity, which are as follows.
\begin{align}
& 0 \leq \phi < 0.37, \quad \quad V_{p1} = (1 - \phi)^2V_{ma} + \phi V_f, \\
& \phi > 0.47, \qquad \qquad \frac{1}{\rho V_{p2}^2} = \frac{\phi}{\rho_f V_f^2} + \frac{1 - \phi}{\rho_{ma}V_{ma}^2}, \\ 
& 0.37 \leq \phi \leq 0.47, \quad \frac{1}{V_p} = \frac{\phi - 0.37}{0.1V_{p2}} + \frac{0.47 - \phi}{0.1V_{p1}},
\label{eq_rhg}
\end{align}
where, $V_{ma}$ and $V_{f}$ denote matrix and fluid velocities, respectively. The data from the two sites has contrasting lithological composition, the KG basin sediment being shaly with around 80-90\% clay content in the rock matrix, and the ME data having low clay content and low porosity values. Due to this contrast, a single model that can fit these lithologies would have to be very nonlinear. 

The goal of the denoising autoencoder for this case would be to receive a suite of logs (4 logs for the examples above), and identify the parts of the data in the logs which do not satisfy the model of \mbox{\cite{Raymer1980}}. One reason could be that one of the 4 logs for those parts of the data is corrupted with noise. Alternatively, it is possible that certain parts of the logs are missing, due to which the check cannot be done. 

To train the denoising autoencoder, a large set of training examples needs to be generated. Prior information related to the possible range of values for $\phi$, $S_{h}$, $V_p$ and $V_{\text{sh}}$ needs to be known. Table \ref{well_table1} lists the ranges for $\phi$, $S_h$ and $V_{\text{sh}}$ that have been used to generate the training set for this example. These ranges correspond to the range of values observed in KG basin and the ME site. In addition, a range of synthetic data values has also been added to make the autoencoder more robust and generalized. From these ranges, random values of $\phi$, $S_h$ and $V_{\text{Sh}}$ are chosen and the corresponding value of $V_p$ is calculated using the model proposed in \cite{Raymer1980}. The log properties are shifted using their respective means and normalized using their maximum and minimum values.  

The training samples need to be fed to the autoencoder network in the form of image slices of size $m \times n$ as shown in Fig. \ref{fig_well_train_image}. Here, $n$ denotes the number of property logs available (equal to 4 for the example considered here). For the $i^{\text{th}}$ data point to be corrected, information from a total of $m$ continuous data points needs to be considered, which includes $\frac{m-1}{2}$ data points from above as well as below the $i^{\text{th}}$ point. In general, well log properties do not vary much between continuous data points. Although we include this fact into our training data, we still allow the properties to vary by up to 20\% between adjacent data points. In this manner, a total of 0.1 million clean images are generated. 

The autoencoder needs corrupted images as well for training. For every clean image, 7 noisy images are generated. To generate a noisy image, one of the 4 logs is randomly chosen, and between 10\% to 40\% data points of this log property are modified. Either up to 10\% random noise is added, or the data at these points is muted. The muted data points are set to -0.1 to differentiate them from the rest of the data. Since there are 4 logs, $n = 4$. Two different autoencoder models are trained with different values of $m$ (3 and 70). The low value 3 is used to understand the local characteristics of the data, and with 70, the goal is to understand the characteristics on a larger scale so that the noisy or muted parts of the data can be differentiated from the noisefree parts. We denote these models by $\bm\Phi_3$ and $\bm\Phi_{70}$.  

\begin{figure*} 
\begin{center}
	\includegraphics{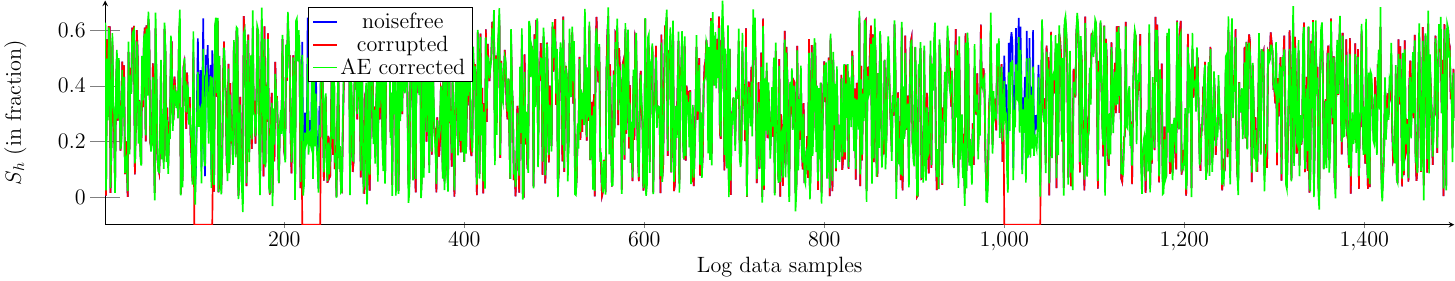}
    \includegraphics{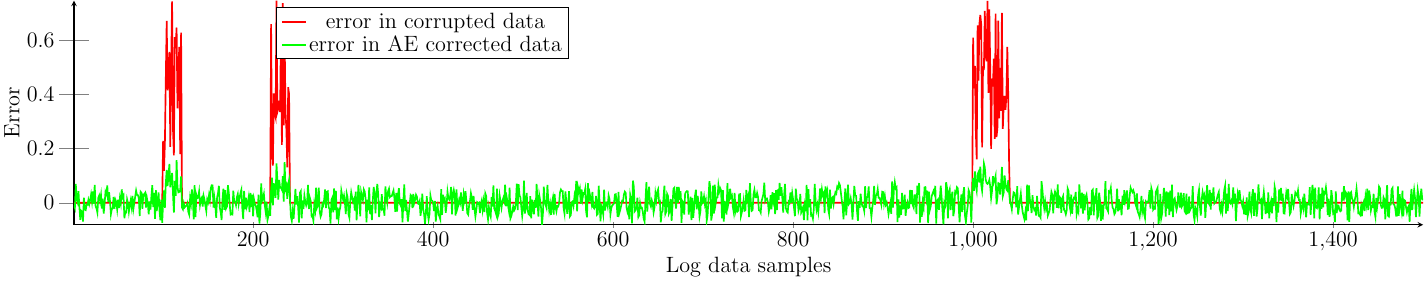}
\end{center}
\vspace{-1em}
\caption{Noisefree, noisy and autoencoder (AE) corrected $S_h$ values for the Mt. Elbert (ME) site, and the error associated with the corrupted and AE corrected logs. The noisy log comprises parts where the data is muted (set to -0.1) and the denoising autoencoder predicts information in these parts.}
\label{fig_rpmresult1}	
\end{figure*}

\begin{figure*} 
\begin{center}
	\includegraphics{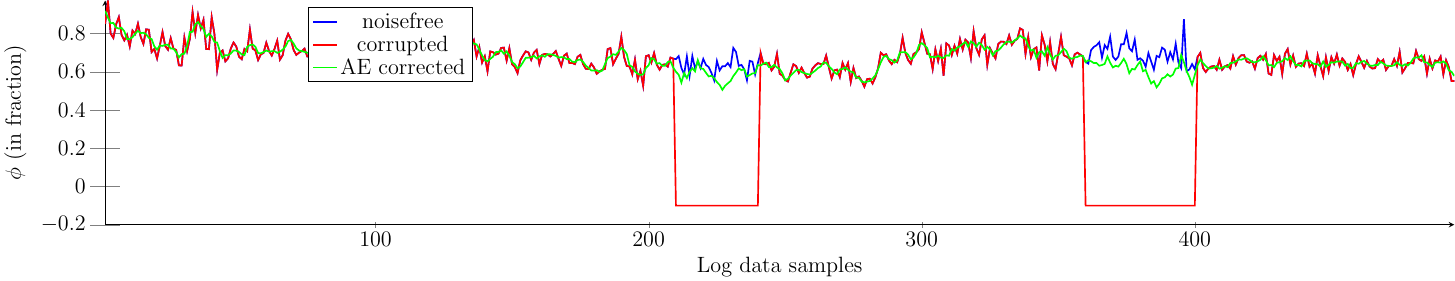}
    \includegraphics{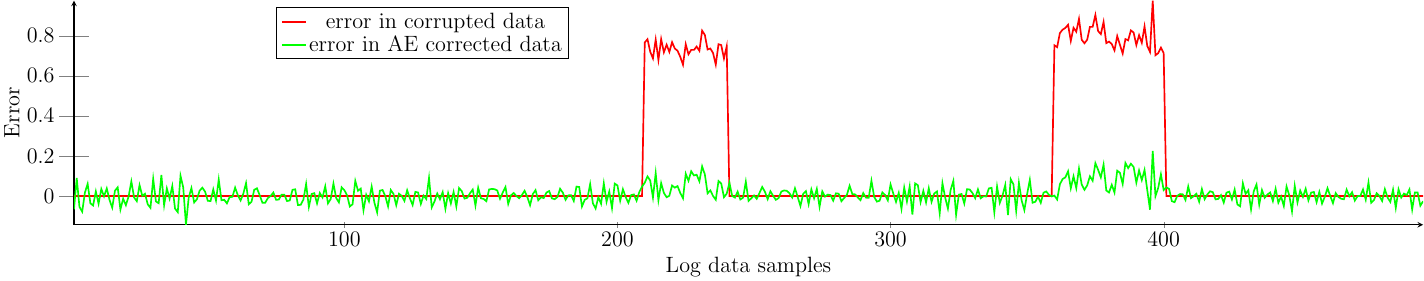}
\end{center}
\vspace{-1em}
\caption{Noisefree, noisy and autoencoder (AE) corrected porosity values $\phi$ for the Krishna-Godavari (KG) basin site, and the error associated with the corrupted and AE corrected logs. The noisy log comprises parts where the data is muted (set to -0.1) and the denoising autoencoder predicts information in these parts.}
\label{fig_rpmresult2}	
\end{figure*}

\begin{figure*} 
\begin{center}
\includegraphics{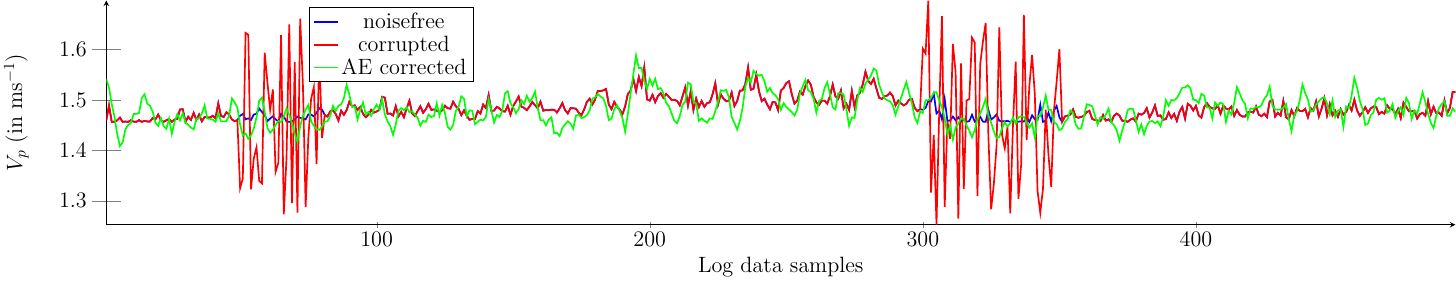}
    \includegraphics{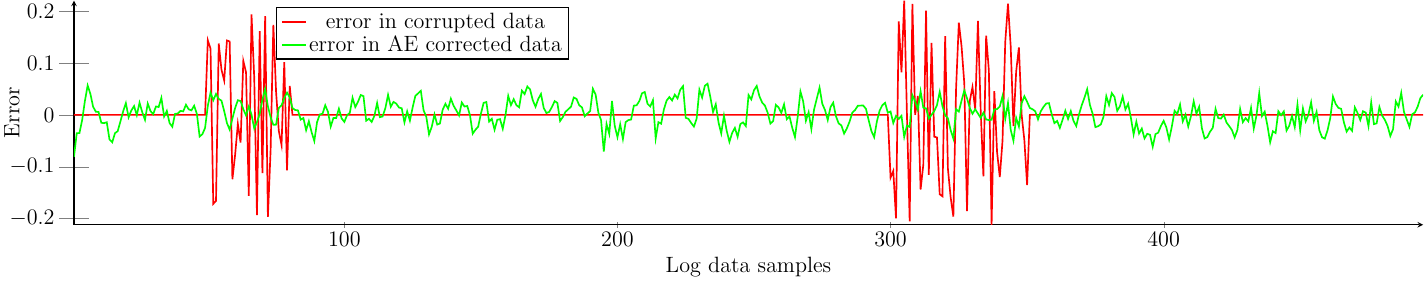}
\end{center}
\vspace{-1em}
\caption{Noisefree, noisy and autoencoder (AE) corrected $V_p$ values for the Krishna-Godavari (KG) basin site, and the error associated with the corrupted and AE corrected logs. The noisy log comprises parts where the random noise has been added to the data and the denoising autoencoder corrects the information in these parts.}
\label{fig_rpmresult3}	
\end{figure*}

The trained autoencoder is then tested on the test set. The real datasets from KG basin and ME site are used for this purpose. Parts of the $S_h$ log from ME site  and $\phi$ log from the KG basin site are muted as shown in Figs. \ref{fig_rpmresult1} and \ref{fig_rpmresult2}. Also, random noise is added in parts of the $V_p$ log of KG basin site as shown in Fig. \ref{fig_rpmresult3}. The objective of the trained autoencoders is to predict the correct information in these parts. For quantitative impression of the added noise, the noisefree data is also shown in Figs. \ref{fig_rpmresult1}, \ref{fig_rpmresult2} and \ref{fig_rpmresult3}. 

From the test dataset, several test image slices are generated as shown in Fig. \ref{fig_well_train_image}. The concept of a sliding window, as discussed in Fig \ref{fig_image_stride}, is used with a stride of 1. Next, the two trained autoencoders $\bm\Phi_3$ and $\bm\Phi_{70}$ are applied on the input test data. The results obtained from the two trained autoencoders are then summed up using weights 0.7 and 0.3, respectively. These weights have been obtained empirically using a trial and error method. 

The final denoised outputs are shown in Figs. \ref{fig_rpmresult1}, \ref{fig_rpmresult2} and \ref{fig_rpmresult3}, respectively. It is observed that the trained autoencoder can nicely predict the missing values of $S_h$ for the ME site and the error in prediction in the missing parts is found to be less than 10\%. For the KG site also, the values of $V_p$ and $\phi$ are predicted very well. However, for the KG site, it is observed that the log data gets regularized to a larger extent, and the resolution is reduced. For all the test examples, some noise gets introduced in the noisefree parts of the data, however, the magnitude of this noise is significantly low. A direction of future research would be to minimize this undesired noise. Nevertheless, from these examples of well data, it is observed that autoencoders can be trained to effectively reduce noise in well log data, as well as predict data in the missing parts of well logs. 

\section{Discussions}
\label{discuss}

In this paper, the applicability of autoencoders has been investigated for the reduction of noise and signal reconstruction in geophysical data. A stacked denoising variant of deep autoencoder network has been proposed, which involves two-step training of the network. Through several numerical examples, we have shown that the proposed methodology works well on geophysical data. However, there are certain limitations of the current methodology, and several research directions can be outlined to improve further on this study. In this section, we briefly look at some of these important aspects.

For better denoising in a deep network regime, we presented a stacked version of deep denoising autoencoders. Note that the presented stacked denoising autoencoder should not be confused with the work of \cite{Vincent2010}, where the stacked denoising formulation aimed at making the autoencoder more robust to noise for classification problems in particular. Moreover, the stacked formulation presented in their work differs significantly from the one presented in this paper. As stated above, during the first step of our stacked denoising autoencoder, the weights of the full network are trained in parts using basic autoencoders with one hidden layer in each. Once the weights are initialized, the second step involves training the full network at once. We have observed that randomly perturbing some of the weights obtained from the first step helps the convergence of the training process in the second step. However, the exact effect of this randomness is unknown yet, and identifying the optimal type of randomness as well as its optimal magnitude is still be to investigated.

Another important aspect that needs to be looked into is the quality check (QC) of the results obtained from the denoising autoencoders. While we confidently show that the proposed autoencoders can significantly reduce noise in the chosen examples, it is difficult to predict how the trained autoencoder will perform on data not represented in the training set. This is a known problem in the field of machine learning, and it would always be advised to limit the application of  a trained network to datasets whose representation overlaps well with the training set. For physics-based problems, similar to the ones considered here, inaccurate removal of noise can change the internal representation of the data, and the whole modeling process can be adversely affected. In this regard, a direction of future research would be to devise a QC approach for machine learning approaches applied on physics-based problems. This QC approach would be expected to provide the extent of uncertainty in the autoencoder output based on the difference between the provided input and the training set. 

The focus of this paper has been restricted to non-coherent noise, and to restrict the length of the paper, we have only considered uniform random noise. Morever, we use mean squared errors which are particularly effective with Gaussian noise. Thus, it needs to be looked into whether choosing a different error function would help to further improve the performance of the denoisers. The application of autoencoders on other non-coherent noise types has been briefly been studied in \cite{Burger2012}. However, in the field of geophysics, removing coherent noise (\emph{e.g.}) ground roll, multiples, \emph{etc}. from seismic data is also a tough challenge. We believe that it would be of interest to the geophysical community to explore the application of stacked denoising autoencoders for the removal of these types of noises as well. Also, the variation of random data that forms an inclusive subset of the training data is limited for the examples considered in this paper. For example, for the seismic test problem considered here, we only assume noisy traces to comprise single frequency component. However, the random noise might have a more complex representation comprising multifrequency components or other functions. Noise based on such representations should also be considered for the robustness of the denoiser.

Moreover, for the seismic problem, we assumed that only a maximum of one trace is corrupt for every small image slice used for training. However, in reality, multiple adjacent traces can be corrupted, and such scenarios should also be modeled. We believe this does not vary the concept demonstrated here, except that the values of $\tilde{m}$ and $\tilde{n}$ chosen for the seismic problem would have to be significantly larger. Due to the availability of limited computational resources, restriction was imposed on the values of $\tilde{m}$ and $\tilde{n}$, however, this would be an interesting direction to look into. Choosing larger image slices allows the autoencoder to interpret a more zoomed-out picture of the representation, thereby providing the capability to interpolate the values of multiple traces at the same time. A similar aspect has been looked into for the well data correction problem considered in this paper.

One limitation observed in the seismic and well data results is that the resolution of the data is compromised during the denoising process. This issue could be suppressed to a certain extent by the use of larger training sets as well as larger training samples in the set. Using a larger training set allows to identify patterns from a wide range of frequencies, which cannot be identified in relatively smaller training sets. Having a wide frequency band plays an important role in improving the resolution of the data. Hence, when larger training set size is used, the resolution of autoencoder output improves. At the same time, it is also important that the low frequency components are preserved so that the zoomed-out representation of the data can be understood. For example, for cases of seismic or well data, where multiple adjacent traces are corrupted, it is important that the pattern of the data is identified on a more global level, and this would require identifying the low frequency representation of the data.

Few additional challenges have been identified that need further investigation for designing improved denoising autoencoders. We observe that in the seismic and well data results obtained from the autoencoder, some noise gets added in the clean parts of the data. This is an undesired noise and needs to be prevented. It is believed that modifying the error (loss) function should help to tackle this issue, and this is still to be investigated. Another aspect is on the choice of continuity in the synthetic well data used for training the autoencoder. In this paper, it is assumed that the the properties between two adjacent samples do not vary by more than 20\%. Based on some preliminary tests, we have observed that the trained autoencoder is very sensitive to the choice of this threshold value. Hence, a direction of research would be to obtain a detailed understanding on the effect of this parameter on autoencoder's performance.

\section{Conclusions}
\label{conclude}
Autoencoders are capable of learning the internal representation of even very complex datasets. Since autoencoders can nonlinearly project data onto a lower dimensional space, several hidden features of the data can be identified, which cannot be realized using the traditional dimensionality reduction techniques. This capability allows identifying the pattern of the signal in the provided data, and separate the noise component. In this paper, the application of autoencoders has been explored in the context of denoising geophysical data. A stacked variant of denoising autoencoders has been formulated, and its application is demonstrated on several numerical examples. For a basic mathematical example, it has been shown that more than 90\% of the random noise can be reduced using denoising autoencoders. For SP anomaly data, the deep networks formulated in this paper could reduce around 80\% of the random noise, when trained using appropriate forward model. The stacked autoencoders are also found to perform very well on seismic and well log data, reducing the random noise and recovering the missing values to a significant extent.

Clearly, the presented stacked denoising autoencoders help to tackle the issue of noise reduction and recovery of missing values in geophysical data. For future work, our goal is to explore the application of these autoencoders on larger datasets, and in the presence of coherent noise. Nevertheless, based on the results presented in this study, it can already be argued that denoising autoencoders could serve as an important data-driven methodology for the elimination of noise in geophysical datasets.

\section*{Acknowledgements}
Parts of the research in this paper have been carried out using \emph{TensorFlow}, an open source software library for high performance numerical computation, especially in the space of machine learning research \cite{Abadi2015}. We would like to thank the developers of this software. Also, we express our thanks to Nikhil Kumar and Jai Gupta for their valuable suggestions and help in the completion of this paper.
\bibliographystyle{spbasic}      

\bibliography{template_new}

\end{document}